\documentclass[aip,apl,reprint,tightenlines,showpacs,superscriptaddress,citeautoscript]{revtex4-1}
\usepackage{bm}
\usepackage{graphicx}

\begin{document}
\title{Full Band Structure Calculation of Two-photon Indirect Absorption in Bulk Silicon}
\author{J. L. Cheng}
\affiliation{Department of Physics and Institute for Optical Sciences,
  University of Toronto, 60 St. George Street, Toronto, Ontario, Canada
  M5S 1A7}
\affiliation{Hefei National Laboratory for Physical Sciences at
  Microscale, University of Science and Technology of China, Hefei,
  Anhui, 230026, China}
\author{J. Rioux}
\affiliation{Department of Physics and Institute for Optical Sciences,
  University of Toronto, 60 St. George Street, Toronto, Ontario, Canada
  M5S 1A7}
\author{J. E. Sipe}
\affiliation{Department of Physics and Institute for Optical Sciences,
  University of Toronto, 60 St. George Street, Toronto, Ontario, Canada
  M5S 1A7}

\date{\today}
\begin{abstract}
Degenerate two-photon indirect absorption in silicon is an important
limiting effect on the use of silicon structures for all-optical
information processing at telecommunication wavelengths. We perform a
full band structure calculation to investigate two-photon indirect absorption in
bulk silicon, using a pseudopotential description of the energy bands
and an adiabatic bond charge model to describe phonon dispersion and
polarization. Our results agree well with some recent experimental
results. The transverse acoustic/optical phonon-assisted processes dominate.
\end{abstract}

\pacs{42.65.-k}
\maketitle

Silicon photonics 
\cite{SiliconPhotonics_Lockwood_Pavesi,IEEE_12_1678_2006_Soref,J.LightwaveTech._23_4222_2005_Lipson}
has attracted widespread interest due to its potential application in
ultracompact all-optical and optical-electronic devices,
particularly for use at telecommunication wavelengths near 1.55~$\mu$m. Many of these
applications are based on the change in index of refraction with
intensity, a third-order optical nonlinearity 
\cite{Opt.Express_215_16604_2007_Lin,Opt.Express_16_1300_2008_Foster,Semi.Sci.Tech._23_064007_2008_Tsang}. Two-photon
absorption, which can plague optical devices because of both
immediate nonlinear loss and subsequent loss due to free-carrier
absorption, is also a third-order nonlinearity. However,
in the interesting wavelength range of 1.2~$\mu$m to 1.7~$\mu$m the sum
of energies from two photons is only sufficient to excite an 
electron across the indirect band gap, with the assistance of an absorbed
or emitted phonon. So indirect degenerate two-photon absorption is of
special interest for the development of optical devices 
\cite{Appl.Phys.Lett._81_1323_2002_Liang,OpticsComm._265_171_2006_Liang,Opt.Express_12_4094_2004_Boyraz,J.Appl.Phys._97_043102_2005_Euser,ElectronicsLett._41_320_2005_Moss}, and
has been the subject of many
investigations \cite{Phys.Rev.Lett._30_901_1973_Reintjes,Appl.Phys.Lett._80_416_2002_Tsang,Appl.Phys.Lett._82_2954_2003_Dinu,Appl.Phys.Lett._90_191104_2007_Bristow,Appl.Phys.Lett._91_071113_2007_Zhang,Appl.Phys.Lett._91_021111_2007_Lin,IEEEJ.QuantumElectron._39_1398_2003_Dinu,J.Phys.B_39_2737_2006_Garcia,Hassan}. 

\begin{figure}[htp]
  \centering
  \includegraphics[height=5.5cm]{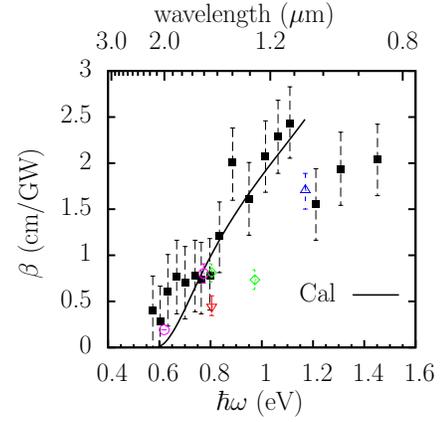}
  \caption{(color online). Spectrum of $\beta$ at 300~K. The circles \cite{J.Appl.Phys._97_043102_2005_Euser},
    uptriangles \cite{Phys.Rev.Lett._30_901_1973_Reintjes}, down 
    triangles \cite{Appl.Phys.Lett._80_416_2002_Tsang}, diamonds
    \cite{Appl.Phys.Lett._82_2954_2003_Dinu}, and squares
    \cite{Appl.Phys.Lett._90_191104_2007_Bristow} are experimental
    results. The solid curve is our calculation.
  }
  \label{fig:compare}
\end{figure}

Experimentally, 
the two-photon absorption coefficient $\beta$ was first measured at
1.06~$\mu$m and 100~K by Reintjes and
McGroddy \cite{Phys.Rev.Lett._30_901_1973_Reintjes}, and found to be 1.5~cm/GW. Thereafter a
series of experiments at telecommunication wavelengths gave $\beta$ in
the range of $0.44-0.9$~cm/GW \cite{Semi.Sci.Tech._23_064007_2008_Tsang}. Bristow
{\it et al.} \cite{Appl.Phys.Lett._90_191104_2007_Bristow} 
measured $\beta$ as a function of photon wavelength and found a strong
dependence (see Fig.~\ref{fig:compare}). Anisotropy in the two-photon absorption was also
investigated \cite{Appl.Phys.Lett._91_071113_2007_Zhang}. Theoretically, most
studies \cite{IEEEJ.QuantumElectron._39_1398_2003_Dinu,J.Phys.B_39_2737_2006_Garcia}
are based on the parabolic band approximation and phenomenological
electron-phonon interactions.
The exciton effect has also been investigated \cite{Hassan},
but only in quantum wells. Yet a full band structure calculation of
the two-photon indirect absorption coefficient is still absent.

In this paper we study the two-photon indirect absorption spectrum in silicon.
We perform a full band structure calculation using an empirical
pseudopotential model (EPM) \cite{Chelikowsky} for the electronic
states, and an adiabatic bond charge model (ABCM) \cite{Phys.Rev.B_15_4789_1977_Weber} for the phonons, and present the
anisotropy and temperature dependence of the absorption. The calculated
two-photon absorption coefficient is in good agreement with some
recent experimental results.

For an incident laser beam with electric field ${\bm E}(t) = {\bm
  E}_\omega e^{-i\omega t} +  c.c$, the rate of the carrier density
injection is written as 
\begin{equation}
  \dot{n} = \xi^{abcd}E_\omega^aE_\omega^b \left(E_\omega^cE_\omega^d\right)^{\ast}\ .
\end{equation}
The superscripts indicate Cartesian coordinates, and repeated superscripts are to be summed
over; $\xi^{abcd}=\sum_{\lambda\pm}\xi^{abcd}_{\lambda\pm}$ are the
injection coefficients, with $\xi^{abcd}_{\lambda\pm}$ standing for
the contribution from the $\lambda^{th}$-branch phonon emission ($+$) or
absorption ($-$) process. From Fermi's Golden Rule we have 
\begin{eqnarray}
  \xi^{abcd}_{\lambda\pm} &=& \frac{2\pi}{\hbar}\sum_{cv\bm
    k_c\in i, \bm k_v}\delta(\varepsilon_{c\bm
  k_c}-\varepsilon_{v\bm k_v}\pm\hbar\Omega^{\lambda}_{\bm
    k_c-\bm k_v}-2\hbar\omega) \nonumber\\
  &&\hspace{-1cm}\times\left[N_{(\bm k_c-\bm k_v)\lambda} +
    \frac{1}{2}\pm\frac{1}{2}\right] W^{ab}_{{c}\bm k_c{v}\bm
      k_v\lambda}\left(W^{cd}_{{c}\bm k_c{v}\bm
      k_v\lambda}\right)^{\ast}\ ,
\label{eq:xi}
\end{eqnarray}
with the transition matrix elements 
\begin{eqnarray}
  W^{ab}_{{c}\bm k_c{v}\bm k_v\lambda} &=&\left\{
\frac{i}{2}\left(\frac{e}{\hbar\omega}\right)^2
  \sum_{{n}{m}}\bigg[\frac{M_{c\bm k_cn\bm k_v;\lambda}v^a_{nm\bm k_v}v^b_{mv\mathbf
      k_v}}{(\omega_{nv\bm k_v}-2\omega)(\omega_{mv\mathbf
      k_v}-\omega)}\right.\nonumber\\
&&\left.\hspace{-1cm}-\frac{v^a_{cn\bm k_c}M_{n\bm k_cm\bm k_v;\lambda}v^b_{mv\bm
    k_v}}{(\omega_{cn\bm k_c}-\omega)(\omega_{mv\bm
    k_v}-\omega)}\right.\nonumber\\
&&\left.\hspace{-1cm}+\frac{v^a_{cn\bm k_c}v^b_{nm\bm k_c}M_{m\bm k_cv\bm
    k_v;\lambda}}{(\omega_{cn\bm k_c}-\omega)(\omega_{cm\bm
    k_c}-2\omega)}\bigg]\right\} + \{a\leftrightarrow b\}\ .
\label{eq:T}
\end{eqnarray}
Here $\bm k_c$ and $\bm k_v$ are the electron and hole wavevectors
respectively; $\{a\leftrightarrow b\}$ stands for the same term as
in the previous $\{\}$, but with the interchange $a\leftrightarrow
b$. The subscript $c$ ($v$) is a band index for
a conduction~(valence) state; $n,m$ identify intermediate
states; $\varepsilon_{n\bm k}$ is the electron energy at band $n$ and
wavevector $\bm k$, and $\hbar\omega_{nm\bm k}=\varepsilon_{n\bm 
  k}-\varepsilon_{m\bm k}$. The phonon energy is given by
$\hbar\Omega_{\bm q\lambda}$ for wavevector $\bm
q$ and branch $\lambda$ (transverse acoustic (TA) and optical (TO),
and longitudinal acoustic (LA) and optical (LO) branches), and $N_{\bm q\lambda}$ is the equilibrium phonon number. The
velocity matrix elements are $\bm v_{nm\bm  k}=\langle n\bm 
k|\frac{\partial H_e}{\partial \bm p}|m\bm k\rangle$, with $H_e$
standing for the single electron Hamiltonian, and the $M_{n\bm k_cm\bm k_v\lambda}=\langle
n\bm k_c|H^{ep}_{\lambda}(\bm k_c-\bm k_v)|m\bm k_v\rangle$ are matrix elements of
the electron-phonon interaction;  $H^{ep}=\sum_{\lambda\bm 
  q}H^{ep}_{\lambda}(\bm q)(a_{\bm q\lambda}+a^{\dag}_{-\bm
  q\lambda})$ is  the electron-phonon interaction, with $a_{\bm
  q\lambda}$ standing for the phonon annihilation operator. 

Due to the $O_h$ symmetry in bulk silicon, the fourth-order tensor
$\xi^{abcd}$ (with $\xi^{abcd}=\xi^{bacd}$ and $\xi^{abcd}=\xi^{abdc}$) only has three independent nonzero components:
\begin{eqnarray}
  && \xi^{xxxx} = \xi^{yyyy} = \xi^{zzzz}\ ,\nonumber\\
  && \xi^{xxyy} = \xi^{xxzz} = \xi^{yyzz} = \xi^{yyxx} = \xi^{zzxx} =
  \xi^{zzyy}\ ,\nonumber\\
  && \xi^{xyxy} = \xi^{xzxz} = \xi^{yzyz}\ ;
\end{eqnarray}
the $\xi^{abcd}_{\lambda\pm}$ share the same symmetry properties. 

For a quantitative calculation, we employ the EPM for the electronic
states and the ABCM for the phonon states; all parameters are
those used in the calculation of one-photon indirect
absorption \cite{longeronephoton}, where good agreement with
experiment has been achieved. Within the pseudopotential scheme we
determine the electron-phonon interaction, and then evaluate the matrix elements
$H_{\lambda}^{ep}(\bm q)$ using the
calculated electron and phonon wavefunctions.
At zero temperature, the calculated indirect band
gap is $E_{ig}=1.17$~eV, but it decreases with increasing temperature \cite{J.Appl.Phys._45_1846_1974_Bludau} 
due to the electron-phonon interaction \cite{Phys.Rev.B_31_2163_1985_Lautenschlager}. Thus
we present our results as a function of the excess photon
energy $2\hbar\omega-E_{ig}$. In the numerical calculation, we use an improved
adaptive linear analytic tetrahedral integration method to perform
the six-fold integration \cite{longeronephoton}, and choose the
lowest 30 bands as intermediate states to ensure convergence. 
According to investigations on one- \cite{Phys.Rev._108_1384_1957_Elliott} and
two-photon \cite{Hassan} indirect gap absorption within the parabolic band
approximation, the neglect of the excitonic effects does not change
the lineshape at energies more than a few binding energies above the
band gap, and we neglect them here. However, since first principle
studies of one-photon direct-gap
absorption \cite{Phys.Rev.B_62_4927_2000_Rohlfing} show that the
electron-hole interaction plays an important role even for higher photon
energies, further work is in order to investigate its role in indirect gap absorption.

We first compare our calculated indirect two-photon absorption coefficient with experiments listed in
Fig.~\ref{fig:compare} for incident light linearly polarized along the
$[001]$ direction; $\beta = 
\frac{2\hbar\omega}{(2n_Rc\epsilon_0)^2}\xi^{xxxx}$. Here $n_R$ is the
refractive index, $c$ is the speed of light, and $\epsilon_0$ is the
vacuum permittivity. We restrict our calculations to photon energies below
the onset of one-photon indirect absorption; beyond this energy
divergences in Eq. (\ref{eq:T}) become problematic and a more
sophisticated theory taking into account real population in
intermediate states is necessary.
Our results are consistent with the measured values of Bristow {\it et
al.} \cite{Appl.Phys.Lett._90_191104_2007_Bristow} The
calculated $\beta$ at $1.55~\mu$m is around
$1.0$~cm/GW. 
\begin{figure}[htp]
  \centering
  \includegraphics[width=7.5cm]{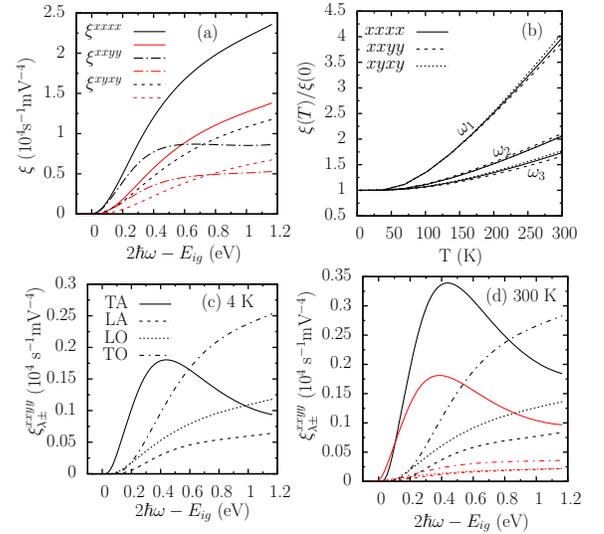}
  \caption{(color online). Detailed calculation results. (a) Spectra
    of $\xi^{xxxx}$, $\xi^{xxyy}$, and $\xi^{xyxy}$ at 4~K (red thin curves) and 300~K (black thick
    curves). (b) Temperature dependence
    of $\xi(T)/\xi(T=0)$ at different photon energies
    $\hbar\omega_i$: $2\hbar\omega_i-E_{ig}=0.1$, $0.43$, and $1$~eV
    for $\omega_1$, $\omega_2$ and $\omega_3$ respectively; at 0~K
    $\omega_2$ corresponds to a wavelength of 1.55~$\mu$m. (c)-(d)
    Phonon-resolved spectra of $\xi^{xxyy}$ at 4~K (c) and 300~K (d);
    black thick (red thin) curves are for
    phonon emission (absorption) processes. }
  \label{fig:detailed}
\end{figure}

When the laser is incident along a direction other than $[001]$, or is
not linearly polarized, other $\xi^{abcd}$ contribute to the
absorption and lead to its anisotropy \cite{Appl.Phys.Lett._91_071113_2007_Zhang}. Fig.~\ref{fig:detailed}
(a) shows the spectra of $\xi^{xxxx}$, $\xi^{xxyy}$, 
and $\xi^{xyxy}$ at temperature 4~K (red thin curves) and
300~K (black thick curves). When the excess photon energy is less than
1.2~eV, all injection coefficients increase with both
photon energy and temperature, except that $\xi^{xxyy}$ increases to a maximum value and then
decreases slightly with photon energy at 300~K. Compared to the one-photon
indirect injection coefficient, which is nearly quadratic in photon
energy \cite{longeronephoton}, the two-photon indirect injection coefficients increase more
slowly due to the explicit dependence on $\omega^{-4}$ and the complicated
frequency dependence of the transition matrix elements appearing in Eq.~(\ref{eq:T}). 

To better understand the spectra, in Fig.~\ref{fig:detailed} we also plot the
phonon-resolved spectra of $\xi_{\lambda+}^{xxyy}$ at 4~K (c) and
300~K (d) as black thick curves. The LA phonon assisted process 
is least important at all temperatures, and gives a contribution less
than 15\%; the TA and TO phonon assisted processes give the most
important contributions, and dominate at low and high photon energies
respectively. The crossover photon energy, where the TO phonon
assisted process first dominates, moves to
higher photon energy with increasing temperature because the TA phonon
has the smaller energy, and the TA phonon assisted process is more
sensitive to temperature. At 300~K, the spectra of
$\xi_{\lambda-}^{xxxx}$ are given as red thin curves in Fig.~\ref{fig:detailed} (d). Only the TA
phonon absorption assisted process gives a significant
contribution. 
Therefore, the photon energy dependence of each absorption process
 is strongly determined by the properties of the mediated phonons, and 
selection rules taking into account the phonon properties are
necessary.

In Fig.~\ref{fig:detailed} (b) we give the temperature dependence of
the injection coefficient at different photon energies. When the temperature is
higher than 200~K, the injection coefficients depend approximately
linearly on temperature, with the slope decreasing with photon
energy. As with one-photon indirect injection \cite{longeronephoton}, the only
temperature dependence comes from the equilibrium phonon number
$N_{\bm q\lambda}$,  which is approximately linear with temperature at
high temperatures. 

In contrast to the two-photon direct absorption coefficient, which has
a cusp \cite{Opt.Mater._3_53_1994_Hutchings,Phys.Rev.B_81_155215_2010_Rioux}
that identifies the onset of absorption from the spin-split off band,
the spectrum of $\beta$ in indirect absorption is very
smooth. This is because the energy dependence at the onset of indirect
absorption is $\propto(\hbar\omega-E_{ig})^2$, 
instead of $\propto(\hbar\omega-E_{g})^{1/2}$ for direct
absorption. In contrast to one-photon indirect gap absorption \cite{longeronephoton}, the TA 
phonon assisted process here is at least as important as TO phonon
assisted process.

In conclusion, we have performed a full band calculation of two-photon
indirect absorption in bulk silicon. We find the two-photon indirect absorption
coefficient is about $1$~cm/GW at $1.55~\mu$m and 300~K. The TA and TO
assisted processes give the most important contributions at all
temperatures. The injection coefficients  depend linearly  on temperature
over 200~K.  

This work was supported by the Natural Sciences and Engineering Research
Council of Canada. J.L.C acknowledges support from China Postdoctoral 
Science Foundation.  J.R. acknowledges support from FQRNT.


\begin{thebibliography}{10}
\bibitem{noteforbeta}
The phonon energy shift is small compared to the energy range we
calculate, and so is neglected.
\end{thebibliography}
\end{document}